\documentclass[twocolumns,conference]{IEEEtran}

\usepackage{url}
\usepackage[pdftex]{graphicx}
\usepackage[cmex10]{amsmath}
\usepackage{amssymb}
\usepackage{amsthm}
\usepackage{float}
\usepackage{epstopdf}
\usepackage{hyperref}
\usepackage{ dsfont,color }
\usepackage{cite}
\usepackage{graphics}
\usepackage{algorithmic}
\usepackage{algorithm}
\usepackage{caption}

\usepackage{bm}

\ifCLASSINFOpdf
\fi
\begin{document}

\title{Sparse Spectrum Sensing in Infrastructure-less Cognitive Radio Networks via Binary Consensus Algorithms\vspace{-1em}}

\author{\IEEEauthorblockN{Mohamed Seif\IEEEauthorrefmark{1}, Tamer ElBatt\IEEEauthorrefmark{1}\IEEEauthorrefmark{2}, and Karim G. Seddik\IEEEauthorrefmark{3}}\\
\IEEEauthorblockA{\IEEEauthorrefmark{1}Wireless Intelligent Networks Center (WINC), Nile University, Egypt\\
\IEEEauthorrefmark{2}Dept. of EECE, Faculty of Engineering, Cairo University, Giza, Egypt\\
\IEEEauthorrefmark{3}Electronics and Communications Engineering Department, American University in Cairo, New Cairo 11835, Egypt\\
        Email: {m.seif@nu.edu.eg,  telbatt@ieee.org,  kseddik@aucegypt.edu}
}}

\maketitle

\begin{abstract}

Compressive Sensing has been utilized in Cognitive Radio Networks (CRNs) to exploit the sparse nature of the occupation of the primary users. Also, distributed spectrum sensing has been proposed to tackle the wireless channel problems, like node or link failures, rather than the common ``centralized approach'' for spectrum sensing. In this paper, we propose a distributed spectrum sensing framework based on consensus algorithms where SU nodes exchange their binary decisions to take global decisions without a fusion center to coordinate the sensing process. Each SU will share its decision with its neighbors, and at every new iteration each SU will take a new decision based on its current decision and the decisions it receives from its neighbors; in the next iteration, each SU will share its new decision with its neighbors. We show via simulations that the detection performance can tend to the performance of majority-rule Fusion Center based CRNs.

\end{abstract}

\IEEEpeerreviewmaketitle





\smallskip
\noindent \textbf{Index Terms:} 
Compressive Sensing, Consensus Algorithms, and Cognitive Radio Network. 
%
\IEEEpeerreviewmaketitle





\section{Introduction}

Cognitive Radio Networks (CRNs) have been highlighted as an elegant solution to the spectrum crunch and to utilize the spectrum under the current spectrum licensing paradigm \cite{fcfcc}. In CRN model, there are two types of users: primary users (PUs) and secondary users (SUs). SUs which do not possess a license to use a spectrum band are nevertheless allowed to transmit when the PUs are sensed inactive. In order to assure a minimal quality of service (QoS) for the PUs, then spectrum sensing is a mandatory task for SUs to detect the presence of the PUs in order to identify the available transmission opportunities \cite{haykin2005cognitive}. Cooperative spectrum sensing has been proposed to combat imperfect conditions of wireless channel environments such as shadowing, fading, and time varying fluctuations in the wireless channel.

Recently, Compressive Sensing (CS) has been investigated as a means for detecting sparsity patterns and recovering sparse signals. CS can reconstruct signal support from a small number of measurements conditioned that the signal is sparse in some domain. With a relation to CRNs, CS has been proposed to reduce the power consumed at the SUs to sense the channels occupied by PUs. Therefore, it is desirable in terms of power consumption to make a small number of linear measurements of the channels instead of scanning each channel seperately \cite{meng2011collaborative}.

In CRNs, the conventional approach for cooperative sensing is the centralized approach. In this sensing paradigm, the measurements of the secondary users are collected by a fusion center, then a final a decision is made based on these reported decisions. However, it is not robust to the wireless channel nature.

Distributed average consensus algorithms \cite{olfati2007consensus} have been investigated in different research areas such as distributed computing, wireless sensor networks, and cooperative control of multi-agent systems. In these problems, the goal is to achieve an average consensus on local information over a network of agents. Average consensus algorithms have been utilized for distributed cooperative spectrum sensing; the authors of \cite{ashrafi2011binary} have proposed two binary consensus approaches: diversity and fusion for infrastructure-less CRNs (i.e., without any aid of fusion center). For diversity based consensus algorithm,  each CR user utilizes its transmissions to repeat its initial vote. After the sensing phase, the algorithm starts in which the CR network is given $K$ time steps (transmissions) to reach a consensus. Each node can use all its transmissions to repeat its initial vote and only fuses the received information at the end of the consensus algorithm. This strategy can, in particular, be useful in reducing the impact of link failures. For the fusion binary consensus approach, each secondary user updates its binary decision, at every step, based on the received votes from its neighbors. In the next time step, it then transmits its updated vote to its neighbors. The convergence of the network performance for fusion-based algorithm is faster than the diversity-based algorithm, however, it is less immune to link failures as the diversity-based algorithm performs.

In this paper, we utilize the tool of CS for infrastructure-less based CRNs where each SU makes linear measurements to detect the occupancy of PUs. Diversity based consensus algorithm is applied in the CRN, we show that the detection performance can tend to the performance of majority-rule Fusion Center (FC) based CRNs depending on the number of algorithm iterations, links' quality between SU nodes, and the number of measurements.

%

The rest of the paper is organized as follows. In Section II, we provide a  mathematical overview on CS and the binary consensus algorithms. The system model is described in Section III. In Section IV, we propose our binary consensus scheme. Simulation results are presented in Section V. Finally, we conclude the paper in Section VI.

\textbf{Notations}: boldface uppercase letters denote matrices and boldface lowercase letters are used for vectors. $\mathds{C}$, $\mathds{R}$ denote the complex and real domain, respectively. For a matrix $\boldsymbol{A}$, its transpose denoted by $\boldsymbol{A}^{T}$.

\section{Preliminaries}

In this section, we briefly review the mathematical background of compressive sensing and binary consensus algorithms.

\subsection{Compressive Sensing}


For an illustration for Compressive Sensing (CS) theory, consider a real-valued signal $\boldsymbol{x} \in \mathds{R}^{N\times 1}$, which is sparse in some domain $\boldsymbol{\Psi}=[\boldsymbol{\psi_{1}}, \boldsymbol{\psi_{2}}, \dots, \boldsymbol{\psi_{N}}] \in \mathds{R^{N \times N}}$. The signal $\boldsymbol{x}$ can be written as follows: 
\begin{equation}
\boldsymbol{x}=\sum_{i=1}^{N}s_{i}\boldsymbol{\psi_{i}}=\boldsymbol{\Psi} s .
\end{equation}
\noindent The signal $\boldsymbol{x}$ is called $K$-sparse if it is represented as a linear combination of only $K$ basis vectors from the domain $\boldsymbol{\Psi}$, that is, only $K$ of the $s_{i}$ coefficients in equation (1) are non zero. 

In CS, only small number of measurements denoted by the vector $\boldsymbol{y}$ is needed to recover the sparse signal $\boldsymbol{x}$ using a $T\times N$ measurement matrix $\boldsymbol{\Phi}$, where $T < N$, as shown in the following equation:
\begin{equation}
\boldsymbol{y}= \boldsymbol{\Phi} \boldsymbol{x} =  \boldsymbol{\Phi} \boldsymbol{\Psi} s = \boldsymbol{\Theta} \boldsymbol{s}
\end{equation}
%

\noindent and the number of required measurements for reliable recovery satisfies the condition $T \geq K \log N$ \cite{baraniuk2007compressive}.

Since $T < N$, the system of equations has no unique solution, and hence, it is impossible to uniquely recover $\boldsymbol{x}$ from $\boldsymbol{y}$. However, given that the vector $\boldsymbol{x}$ is $K$-sparse, where $K << N$, then the CS theory allows us to reconstruct the signal, provided that the measurement matrix $\boldsymbol{\Phi}$ is chosen such that it satisfies the Restricted Isometric Property (RIP) condition, as follows.
\begin{equation}
(1-\delta) \left\|\boldsymbol{x}\right\| ^{2}_{2}  \leq \left\|\boldsymbol{\Phi} \boldsymbol{x} \right\|^{2} _{2} \leq (1+\delta)  \left\|\boldsymbol{x} \right\|^{2} _{2}.
\end{equation}

The signal reconstruction problem can be formulated as an $\ell_1$-norm minimization problem as follows.
\begin{equation}
\min_{\hat{\boldsymbol{x}} \in \mathds{R}^{N}} \left\| \hat{\boldsymbol{x}} \right\| _{1} \hspace{0.3in} s.t. \left\|\boldsymbol{y}- \boldsymbol{\Phi} \hat{\boldsymbol{x}} \right\| _{2} \leq \epsilon
\end{equation}
\noindent where $\epsilon$ is some small tolerance dependent on the noise level. Different recovery techniques were developed to improve the computational complexity of compressive sensing reconstruction. Techniques such as ``subspace pursuit'' and ``orthogonal matching pursuit'' have been developed that exhibit computational complexity of  $\mathcal{O}(NM)$ and $\mathcal{O}(N \log M)$, respectively.  Also the authors in \cite{meng2011collaborative} proposed two compressive sensing based collaborative spectrum sensing techniques that are computationally efficient; the first approach is using matrix completion technique in which the technique reconstructs a low rank matrix from small number of entries. The second approach is based on joint sparsity recovery. This approach utilizes the correlation between sparsity patterns measured by the CRs, however, it is mentioned that the correlation between the measurements will be reduced in the presence of noise and wireless channel environments.

\subsection{Binary Consensus Algorithm}
In this subsection, we provide a basic review of binary consensus algorithms for distributed spectrum sensing\cite{olfati2007consensus}. To allow SUs to cooperatively arrive at a global decision and with no help from a designated central entity, each SU makes a local decision regarding the presence or absence of a PU, denoted as $H_1$ or $H_0$, respectively. SUs then exchange their binary local decisions with their direct neighbors for $K$ time steps, where $K$ is the running time of the algorithm. Upon the termination of the algorithm, each SU individually makes a decision of $H_1$ or $H_0$, based on the received decisions from neighboring nodes. Let $\boldsymbol{b}(k)=[b_{1}(k),\dots,b_{M}(k)]^{T}$ be the vector of local decisions at time step $k$ at the $M$ SUs. The binary consensus algorithm can be summarized as follows.
\begin{enumerate}
 \item At the first time step ($k=0$), each SU initially transmits its local decision to its neighbors that are connected to it at this time step.
 \item At each consecutive time step ($0 < k < K$), each SU collects the decisions transmitted by its neighboring SUs. It then combines these decisions, along with the previously received decisions from past time steps, through a \textit{combining function} which generates a new decision $\boldsymbol{b}(k)$; this new decision is to be transmitted to neighboring nodes at the current time step $k$. This can be mathematically expressed as:
 \begin{equation}
 \label{combining_function}
  \boldsymbol{b}(k) = \mathcal{F}(\boldsymbol{b}(n), \; n=0,\cdots,k-1), \hspace{0.05in} 0<k<K-1
 \end{equation}
 \noindent where $\mathcal{F}(.)$ is the combining function.
 \item Upon the termination of the algorithm ($k=K$), each node makes a final decision based on the previously obtained decisions from all time steps $0<k<K$, through a \textit{decision function}. This can be mathematically expressed as:
  \begin{equation}
 \label{decision_function}
  \boldsymbol{b}(K) = \mathcal{D}(\boldsymbol{b}(n), \; n=0,\cdots,K-1)
 \end{equation}
  \noindent where $\mathcal{D}(.)$ is the decision function.
\end{enumerate}

Based on the appropriate choice of the combining and decision functions, the binary consensus algorithm is guaranteed to converge to a common decision after a sufficiently long running time, i.e., $b_{i}(K) = b^\star, \; \forall i=1,\cdots,M$ as $K \rightarrow \infty$\cite{olfati2007consensus}. 

In this paper, we focus on one variation of binary consensus algorithms, namely, \textit{diversity-based binary consensus algorithm}\cite{ashrafi2011binary}, in which a SU uses its initial local decision for decision reporting at all consecutive time steps, and the combining function is basically a majority rule for the received decisions along the $K$ time steps. The combining and decision functions in this case are mathematically expressed as:

\vspace{- 0.1 in} 

\begin{equation}
\begin{split}
 \mathcal{F}(\boldsymbol{b}(n), \; n=0,\cdots,&k-1) = \boldsymbol{b}(k-1), \hspace{20pt} 1< k <K, \\
 \mathcal{D}(\boldsymbol{b}(n), \; n=0,\cdots,&K-1) = \\
 &\text{Dec}\bigg(\dfrac{1}{M}(\boldsymbol{b}(0)+\dfrac{1}{Kp}\sum\limits_{t=0}^{K-1} \boldsymbol{A}(t)b(t))\bigg)
\end{split}
\end{equation}

\noindent where $\text{Dec}(x)=\begin{cases}
                                      1, & \text{if } x \geq 0.5\\
                                      0, & \text{if } x < 0.5
                                     \end{cases}$
\noindent with 1 and 0 corresponding to deciding $H_1$ and $H_0$, respectively. When $x$ is a vector, the function operates on it element-wise.

It is proven in \cite{ashrafi2011binary} that the global probability of detection can be expressed as follows  
 
\begin{eqnarray}
P_{d}(K) &\approx & \sum_{i=0}^{M} \left( \begin{matrix} M \\ i \end{matrix} \right) \left[ (1-\pi_{11}) \hspace{-0.15in}\quad Q\bigg(\frac { (\frac{M}{2} - i) \sqrt{K}}{ \sqrt{\frac{1-p}{p}i} } \bigg) \right]^{M-i}\\ \nonumber
&\times &   \left[ \pi_{11}\hspace{-0.15in}\quad Q\bigg(\frac { (\frac{M}{2} - i) \sqrt{K}}{ \sqrt{\frac{1-p}{p}|i-1|} } \bigg) \right]^{i}
\end{eqnarray}

\noindent where $Q(x)= \frac{1}{\sqrt{2\pi}}\int_{x}^{\infty} e^{-t^{2}/2} \text{d}t$ and $\pi_{11}$ is the probability of detection at the $i{\text{th}}$ SU.

The asymptotic behaviour the algorithm can be written as

\begin{equation}
\lim_{K \rightarrow \infty}P_{d}(K)= \sum_{i=\lceil \frac{M}{2}\rceil}^{M} \left( \begin{matrix} M \\ i \end{matrix} \right) (1-\pi_{11})^{M-i} \pi_{11}^{i}
\end{equation} 

\noindent which implies that the performance of the binary consensus algorithm will converge to the performance of the majority decision rule as $K \rightarrow \infty$. It is worth noting that, the asymptotic behaviour of the diversity-based consensus algorithm is independent of the network connectivity and depends on the number of SUs and signal to signal-to-noise ratio (SNR). Moreover, the false alarm probability is independent of signal power and link quality, and it only  depends on the noise parameter.


\subsubsection*{\textbf{Vector consensus}} In this work, we extend the consensus on a scalar value, i.e., PU occupancy of single channel into an approximate consensus on vector, i.e., the PUs occupancy of multiple channels. Consequently, this consensus problem maps to a vector consensus problem. Approximate vector consensus is characterized by two important things: 1) $\epsilon$-agreement which indicates how far the output vector of each node is from the true vector, 2) Termination, in which the approximate consensus (i.e., the difference between each element in the output vector and the true vector does not exceed $\epsilon$) can be reached in finite time \cite{vaidya2014iterative}.


 \section{System Model}

 
 \subsection*{Signal Model}
We consider a cognitive radio network with $M$ SU nodes that locally monitor a subset of $N$ channels. A channel is either occupied by a PU node or unoccupied, corresponding to the states $1$ and $0$, respectively. We assume that the number $P$ of occupied channels is much smaller than $N$. The goal is to recover the occupied channels from the SU nodes observations. Since each SU can only sense limited spectrum at a time, it is impossible for limited $M$ SUs to observe $N$ channels simultaneously. Therefore, each SUs node makes a small number of measurements $T$ using a measurement matrix $\textbf{F}_{i}$ whose elements are drawn from a Gaussian distribution \cite{meng2011collaborative}. 

\begin{figure*}
 \begin{minipage}{0.3\textwidth}
  \includegraphics[scale=0.35]{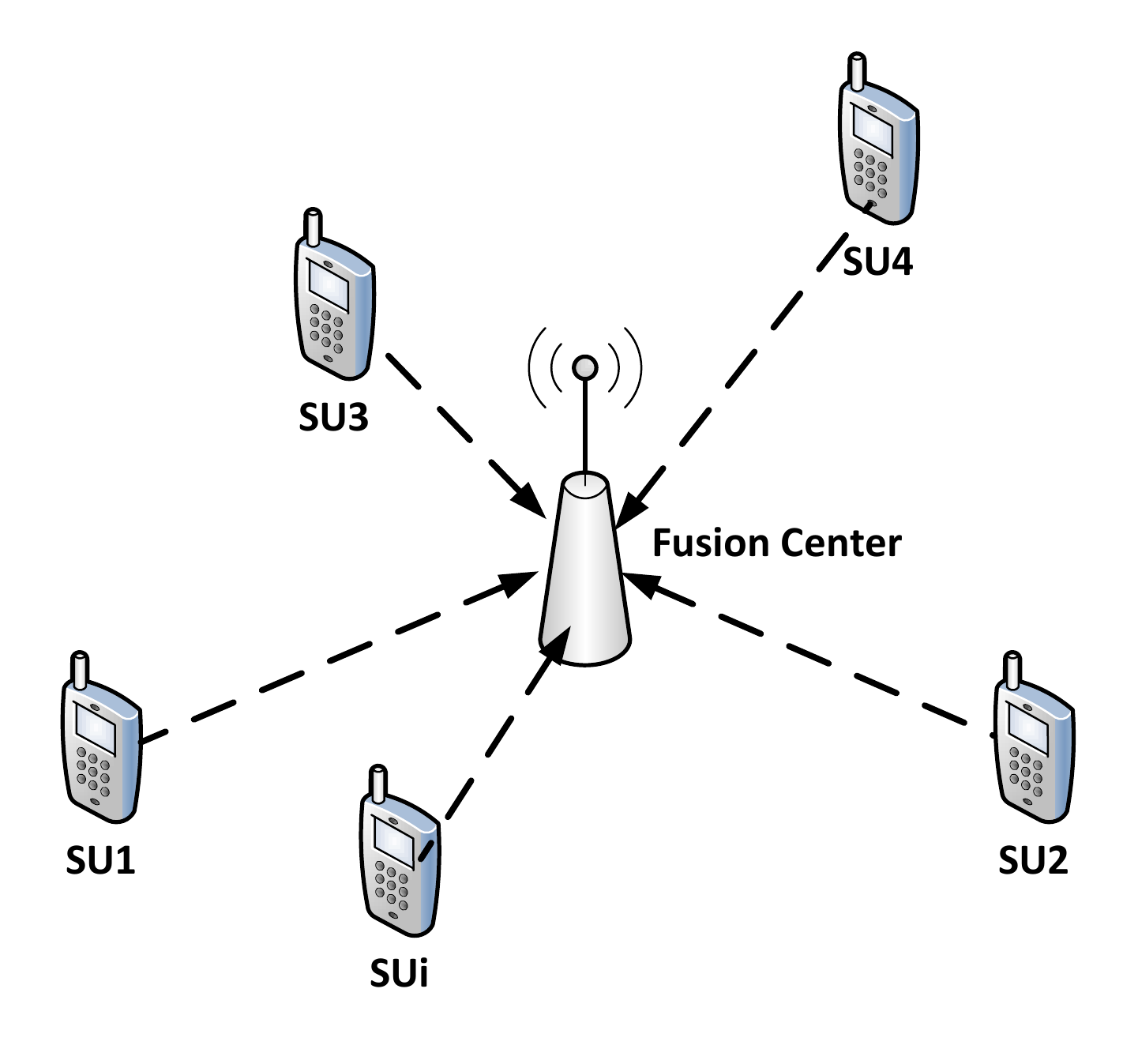}
\caption{Fusion-based CRN.}
 \end{minipage}
 \hspace{1.3in}
 \centering
 \begin{minipage}{0.3\textwidth}
   \includegraphics[scale=0.35]{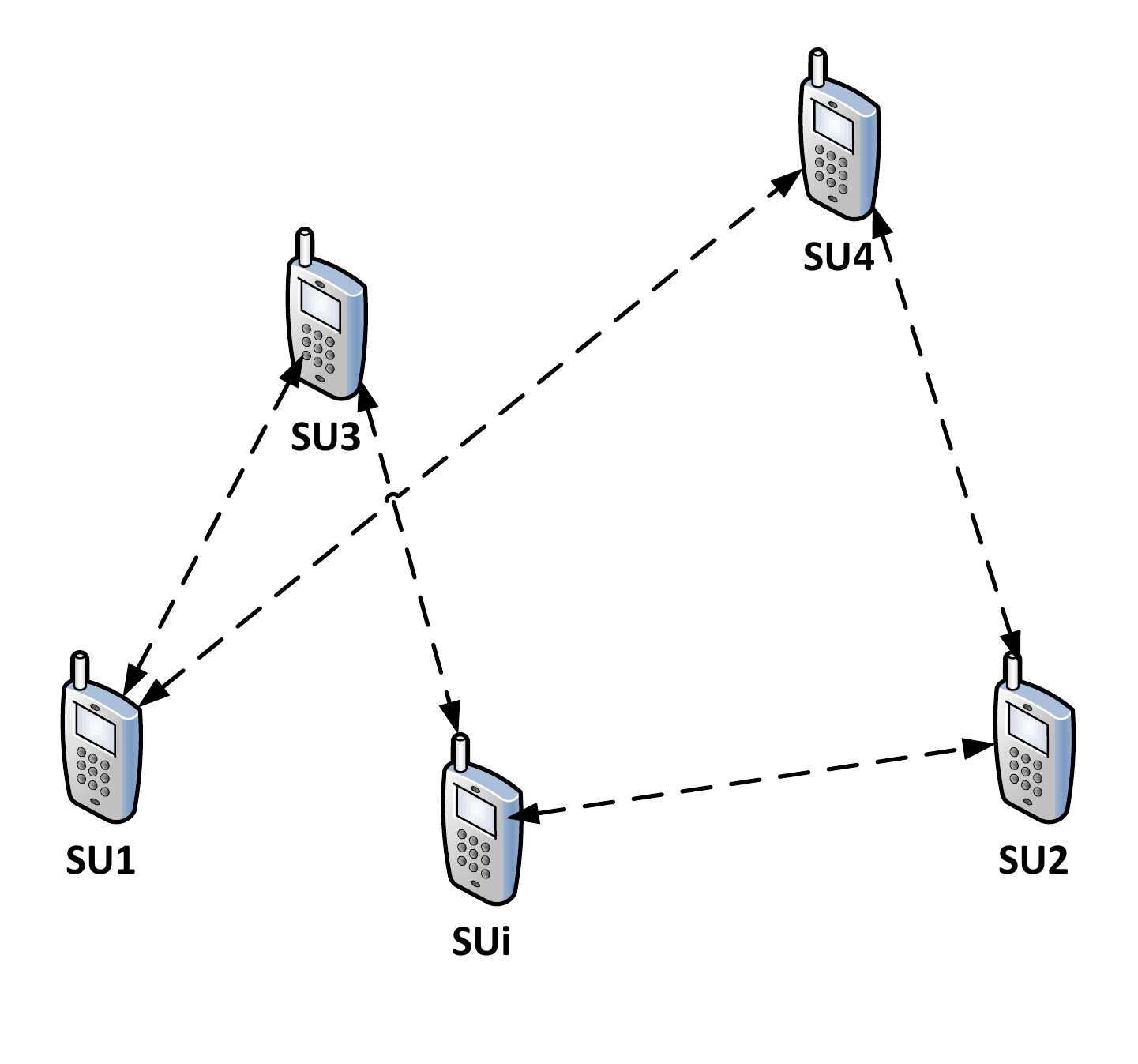}
\caption{Infrastructure-less CRN.}
 \end{minipage}
\end{figure*}

 The channel gain $G_{i,j}$ between the $i$th SU and the channel $j$ is defined as follows:
\begin{equation}
G_{i,j}=d_{i,j}^{-\alpha/2} \left|h_{i,j}\right|,
\end{equation} 

\noindent where $d_{i,j}$ is the distance between the $i$th SU and the $j$th PU, $\alpha$ is the path loss exponent, and $h_{i,j}$ is a complex-Gaussian channel gain between the $i$th SU and the $j$th PU; without loss of generality, the transmitted signal from the $j$th PU is normalized to unity.

 \begin{figure}[ht!]
        \includegraphics[scale=0.38]{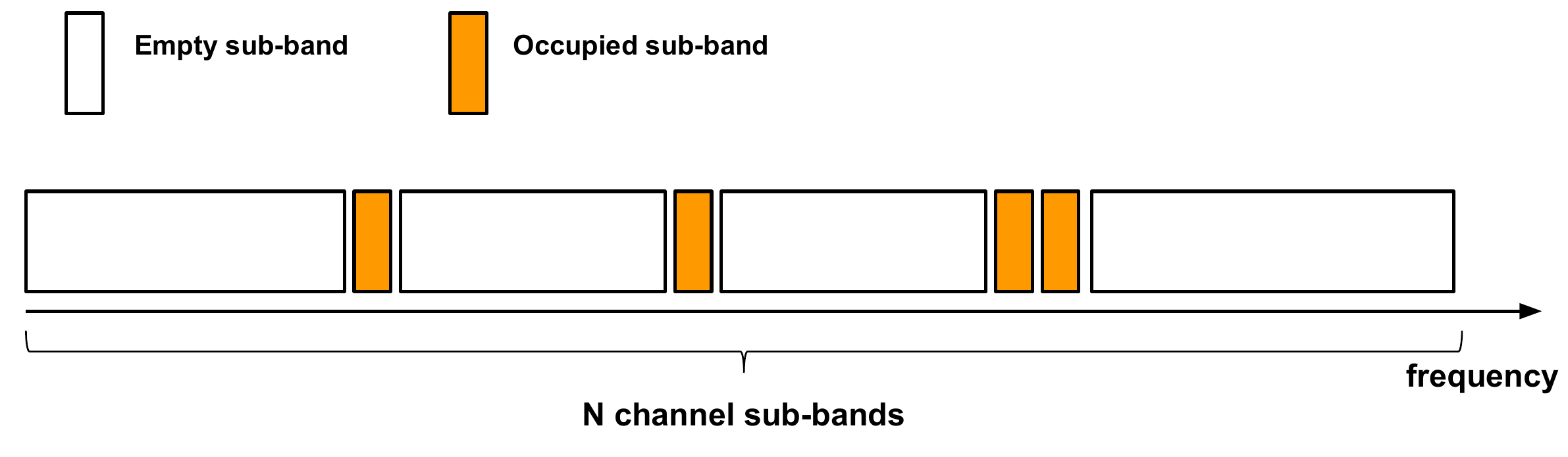}
        \caption{Sparsity Nature of Spectrum Occupation by PUs.}
     \end{figure}

Collectively, the PUs signals of the network can be written as follows: 
\begin{equation}
\boldsymbol{X}_{N \times M} = \boldsymbol{R}_{N \times N } \times (\boldsymbol{G}_{M \times N})^{T}
\end{equation}
\noindent where $\boldsymbol{R} \in \mathds{R}^{N \times N}$ is a diagonal matrix, whose diagonal elements represent the states of all channels whether active '1' or inactive '0', and $\boldsymbol{G} \in \boldsymbol{C}^{M \times N}$ is the channel gain matrix between the PUs and SUs whose its elements $G_{i,j}$. 

\noindent Also, the received signal can be written as: 
\begin{equation}
\boldsymbol{Y}_{T \times M}=\boldsymbol{F}_{T \times N} \boldsymbol{X}_{N \times M} + \boldsymbol{W}_{T \times M}
\end{equation}

\noindent where is $\boldsymbol{W}$ is the measurement noise matrix, whose elements are independent with zero mean and a variance of $\sigma^{2}$.

\section{Proposed Sensing Scheme}

\subsection*{Phase one:  Independent $\ell_{1}$-norm Recovery}
After each SU node $i$  makes its measurements, each SU $i$ applies the following $\ell_{1}$-norm  recovery algorithm \cite{candes2006robust}:\\
\begin{equation}
 \min_{\hat{\boldsymbol{x}}_{j} \in \mathds{R}^{N}} \left\| \hat{\boldsymbol{x}}_{j} \right\| _{1} \hspace{0.3in} 
\end{equation}
\begin{equation}
s.t. \left\|\boldsymbol{F}_{j} \hat{\boldsymbol{x}}_{j}-\boldsymbol{y}_{j} \right\| _{2}^{2} \leq \sigma_{j}^{2}, \hspace{0.1in}  \forall j \in \{1, \dots, M\}
\end{equation}

\noindent where $\hat{\textbf{x}}_{j}$ is the vector of recovered signal at SU $j$, and $\sigma_{i}^{2}$ is estimated noise power.

Then, we employ a threshold test at each SU node to generate the decision vector $\boldsymbol{b}_{j}(0), \forall j\in \{1, \dots, M\}$ about the occupancy of the $N$ channels. 
\begin{equation}
 b_{ji}(0) = \left\{
 \begin{array}{ll}
            1,  & \hbox{$x_{ji}(0) \geq \eta$} \\
            0, & \hbox{$ x_{ji}(0) < \eta$} \\
            \end{array}\right.
\end{equation}
\noindent where $b_{ji}(0)$ is the initial binary decision about the $j$th channel at the $i$th SU node.

\subsection*{Phase two: Consensus Algorithm}

\subsubsection*{\textbf{Network model between the CR nodes}}
 We model the secondary network as an undirected random graph $\mathcal{G} (M, E)$,
where $M$, the set of nodes, represents the SUs, and $E$, the set of edges, denotes the connectivity of SUs. A node $i$ is connected to node $j$ if $\tau_{ij}(k) > \tau$, where $\tau_{ij}(k)$ is the instantaneous SNR of the signal of SU $i$ at SU $j$ at time step index $k$, and $\tau$ is the minimum acceptable SNR required for successful decoding of secondary transmission. Assuming channel reciprocity, then $\tau_{ji}(k) = \tau_{ij}(k)$ and both SUs are in the neighborhood of each other if their instantaneous SNR exceeds the decoding threshold. The probabilistic nature of the wireless channel and consequently, of the instantaneous SNR of received secondary signals are the reasons behind the ``randomness'' in the network graph. Due to the absence of a central entity for coordinating transmissions,
nodes which are in the same transmission range of each other exchange their decisions.

\begin{table}
\footnotesize
 \begin{tabular}{|c|l|}
 \hline
  \textbf{Symbol} &\textbf{Description} \\
  \hline
  $M$ & Number of SUs. \\
  $\bar{\tau}_{ij}(k)$ & Instantaneous SNR of $i$th SU at $j$th SU at time step k. \\
  $\tau$ & Min. acceptable threshold for successful decoding. \\
  $\alpha$ & Path loss exponent.\\
  $T$ & The number of measurements at each SU node. \\
  $\textbf{F}_{i}$ & The measurement matrix of the $i$th SU.\\
  $\textbf{b}_i(0)$ & Initial binary decision vector of $i$th SU. \\
  $N$ & Number of frequency sub bands.\\
  $\textbf{A}$ & Adjacency matrix\\
  $\eta$ & Local detection threshold. \\
  \hline
 \end{tabular}
\label{table::list_symbols}
\caption{List of symbols.}
\end{table}

To model the connectivity between SUs in the network, the \textit{adjacency matrix} $\boldsymbol{A}(k) \in \mathds{R}^{M \times M}$ is defined as:
\begin{equation}
a_{ij}(k) = \begin{cases}
1 &  \text{if $\bar{\tau}_{ij}(k) >= \tau, \: i \neq j$}  \\
0 & \text{otherwise}
\end{cases}
\end{equation}
\noindent where $a_{ij}(k), \; i,j \in \{1, \; \cdots \;,M \}$ denotes the $(i,j)$th element of the matrix $\boldsymbol{A}(k)$. $a_{ij}(k)=1$ in this context means that nodes $i$ and $j$ are connected. For ease of exposition, we neglect the intricate details of the wireless channel transmission and communication scheme and assume $a_{ij}(k), \; i \neq j, \; \forall \; k \geq 0$ to be Bernoulli random processes with $p=\text{Pr}(a_{ij}(k)=1)=\text{Pr}(\bar{\tau}_{ji}(k)>\tau)$, which implicitly models wireless channel characteristics.

We apply the following consensus scheme for $K$ iterations between the SU nodes:
\begin{equation}
\boldsymbol{b}_{j}(k) =\text{Dec}\bigg(\dfrac{1}{M}(\boldsymbol{b}(0)+\dfrac{1}{Kp}\sum\limits_{t=1}^{K} \boldsymbol{B}(t)\boldsymbol{a}_{j}^{T}(t))\bigg)
\end{equation}
\noindent where $\boldsymbol{b}_{j}(k)$ is the decision vector at the $j$th SU node at time $k$,  $\boldsymbol{B}(t)=\left[\boldsymbol{b}_{1}(t), \boldsymbol{b}_{2}(t), \dots, \boldsymbol{b}_{M}(t)\right] \in \mathds{R}^{N \times M}$, and $\boldsymbol{a}_{j} \in \mathds{R}^{M}$ is the $j$th column vector of the adjacency matrix $\boldsymbol{A}$. 

\subsubsection*{\textbf{Convergence of the consensus algorithm}}

\noindent The updated term in the algorithm at a certain time instant can be written as (for convenience, we remove the time index) 
\begin{equation}
\boldsymbol{u_{j}}=\boldsymbol{B} \boldsymbol{a}_{j}^{T}=\left[ \begin{matrix} { b }_{ 11 } & \cdots  & { b }_{ 1M } \\ \vdots  & \ddots  & \vdots  \\ { b }_{ N1 } & \cdots  & { b }_{ NM } \end{matrix} \right] \left[ \begin{matrix} { a }_{ j1 } \\ \vdots  \\ { a }_{ jM } \end{matrix} \right].
\end{equation}
\noindent Then, for channel $j$, the updated term at the $i$th SU will be as follows
\begin{equation}
u_{ji}= b_{j1}a_{i1}+b_{j2}a_{i2}+\dots+\underbrace { b_{ ji }a_{ ii } }_{ 0 } +\dots+b_{jM}a_{iM}  .
\end{equation}
\noindent Using lemma 1 in \cite{ashrafi2011binary}, we can show that
\begin{equation}
\lim_{K \rightarrow \infty} \text{Dec}\frac{1}{M}(\bigg(b_{ji}(0)+\frac{1}{Kp}\sum _{ t=1 }^{ K }{ { u }_{ ji }(t) })\bigg) = b_{j}^{*}, 
\end{equation}
\begin{equation*}
\forall j= 1,\dots, N, \hspace{0.1in} \forall i = 1, \dots, M.
\end{equation*}
Then, the convergence is held for each element of the decision vector $\boldsymbol{b}_{i}$, $\forall i=1,\dots, M $. Then, it can be verified that the detection probability will converge to that of the majority-rule based fusion system:
\begin{equation}
\lim_{K \rightarrow \infty}P_{d}(K)= \sum_{j=1}^{N}\sum_{i=\lceil \frac{M}{2}\rceil}^{M} \left( \begin{matrix} M \\ i \end{matrix} \right) (1-\pi_{11|R_{jj}=1})^{M-i} \pi_{11|R_{jj}=1}^{i}.
\end{equation} 

\section{Simulation Results}

In this section, we investigate the performance of our proposed algorithm through simulations. We compute the probability of detection at time $k$ as follows:
\begin{equation}
P_{d}(k)=\frac{\sum_{j | R_{jj}=1} \sum_{i = 1}^{M} \mathds{1}
(b_{ji}(k) = 1 | R_{jj}=1)}{M \times P}
\end{equation}
\noindent and for the false alarm probability, we calculate it as
\begin{equation}
P_{fa}(k)=\frac{\sum_{j | R_{jj}=1} \sum_{i=1}^{M}  \mathds{1}
(b_{ji}(k)= 1| R_{jj}=0)}{M \times (N-P)}
\end{equation}
\noindent where $\mathds{1}(A)$ is the indicator function for an event $A$ and takes the value '1' if $A$ is valid and '0' otherwise. In addition, we compare the performance of the proposed distributed algorithm with a ``centralized'' fusion algorithm as an upper bound for the infrastructure-less based performance at different SNRs. For the fusion based CRN\footnote{The link quality for the fusion based system is assumed to be perfect; $p=1$.  } we use the majority-rule decision making, for which the probability of detection is given as follows:\\
\begin{equation}
P_{d}=\frac{\sum_{j | R_{jj}=1} \mathds{1}
(D_{j} >=  M/2  | R_{jj}=1)}{P}
\end{equation}
\noindent and for the false alarm probability is given by
\begin{equation}
P_{fa}=\frac{\sum_{j | R_{jj}=0} \mathds{1}
(D_{j} >= M/2  | R_{jj}=1)}{N-P}
\end{equation}
\noindent where $D_{j} = \sum_{i=1}^{M} b_{ji}$. \\


 The simulated model consists of $N = 200$ channels. The locations of PUs are uniformly distributed over an area of $1000 \times 1000$ $\text{m}^{2}$, with a minimum distance of $10$ m between any two SUs, and the pathloss exponent is set to be $\alpha=2$. We set the algorithm iterations number to be $K = 10$. We have randomized the simulation parameters for $500$ trials 

\begin{figure}[ht!]
       \includegraphics[scale=0.6]{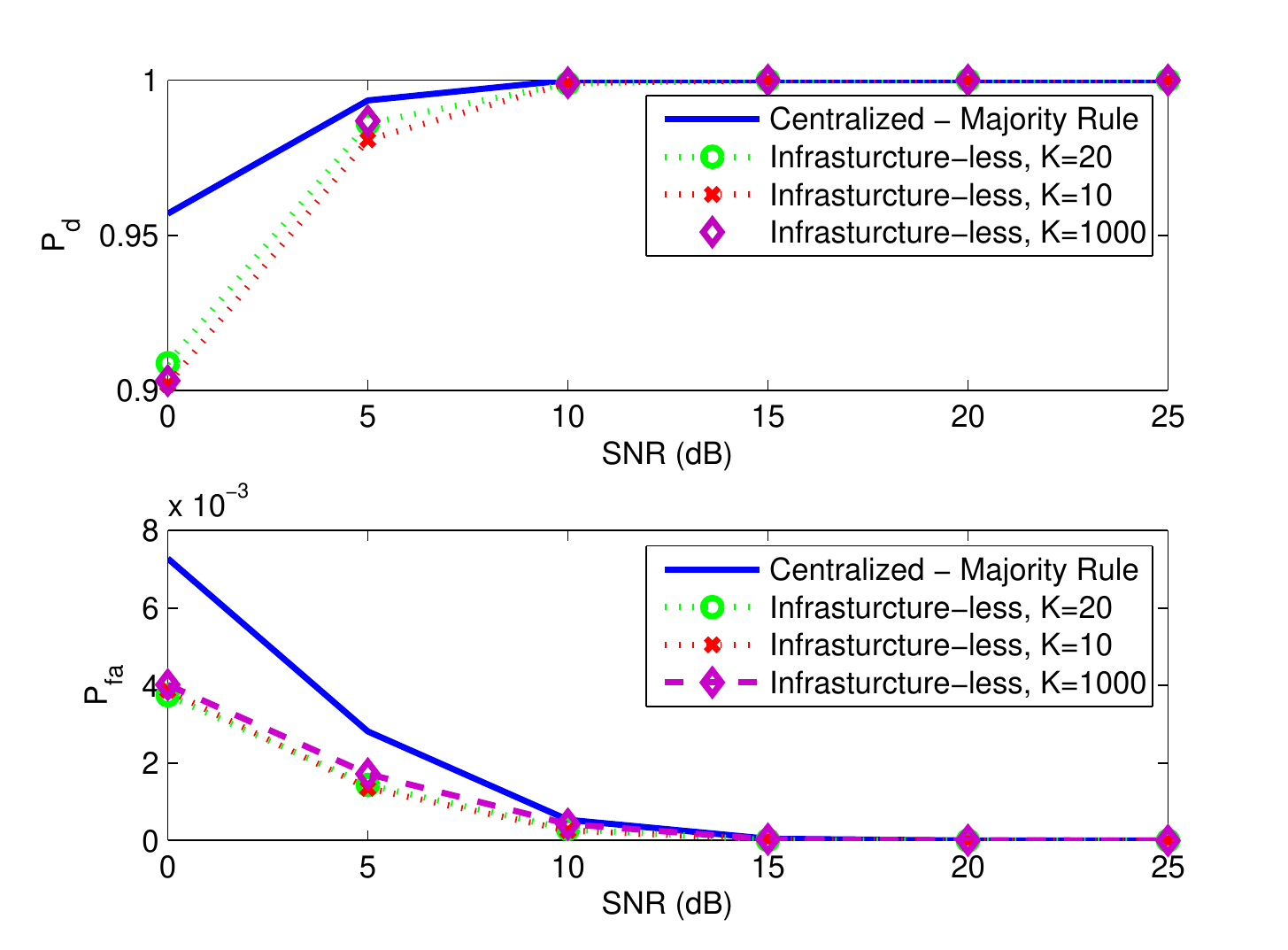}
        \caption{\small{Performance comparison between fusion based CRN and infrastructure-less CRN: N = 200 channels, T = 50 measurements, P = 4 users, M = 12 users, $\alpha$ = 2. }}
        \label{fig::Comparsion}
     \end{figure}

\begin{figure}[ht!]
       \includegraphics[width=90mm, height=60mm]{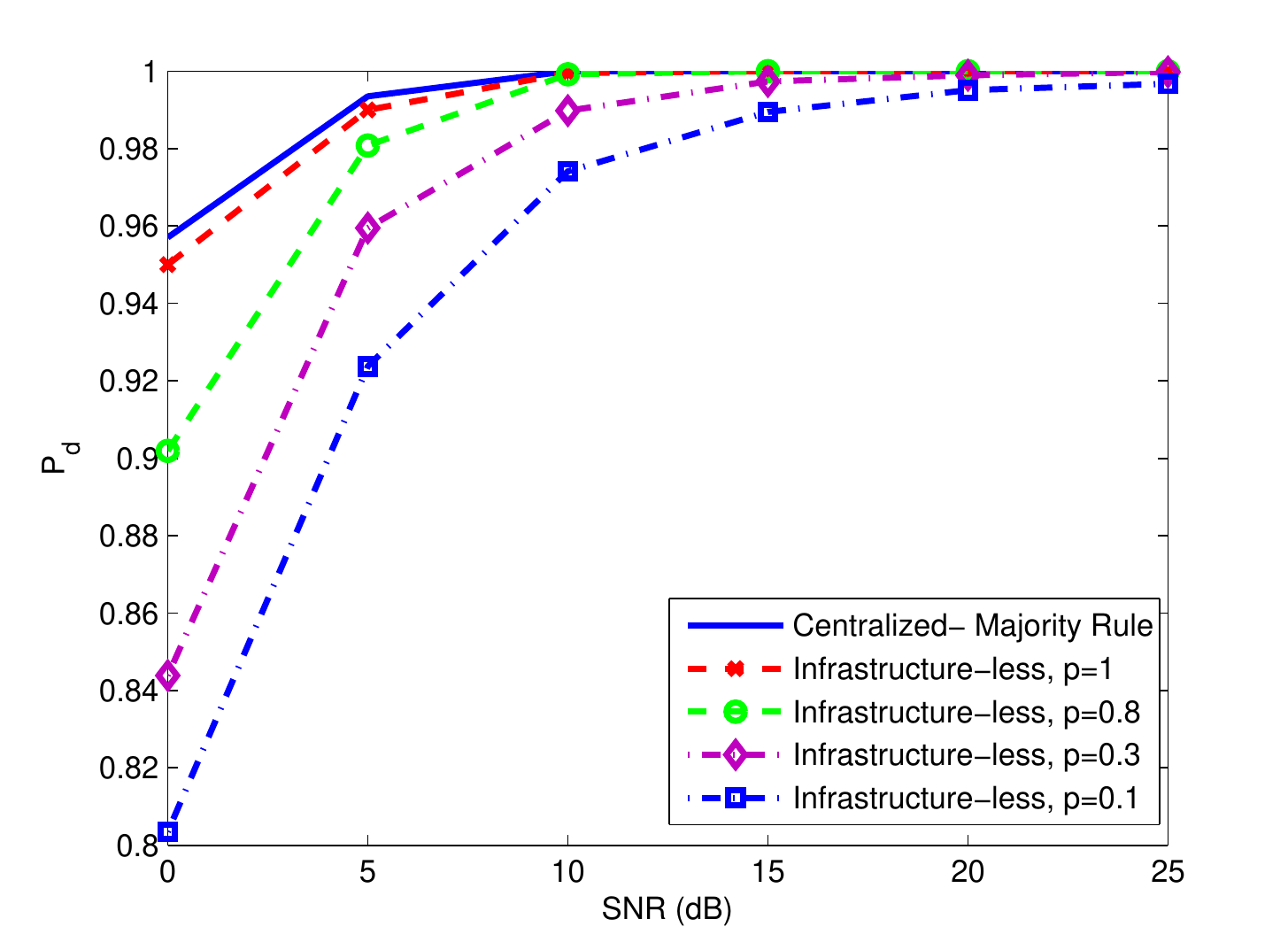}
       \caption{\small{Effect of link quality on probability of detection: N = 200 channels, T = 50 measurements, P = 4 users, M = 12 users, $\alpha$ = 2.} }
       \label{fig::Link}
    \end{figure}

Fig. \ref{fig::Comparsion} shows the effect of the number of iterations on the network performance, i.e., the asymptotic behaviour of the algorithm. It is clear that the performance of the algorithm will converge to a constant value after nearly $K = 20$. Another observation is that the performance of infrastructure-less network is upper bounded by the performance of the centralized network, which, in part, is due to the fact that not every SU is connected to all the other SUs. Fig. \ref{fig::Link} shows the effect of the link connectivity on the detection performance. It is clear that as the probability to establish links between the SU nodes increases, the performance of the detection algorithm enhances in terms of both the detection probability and the false-alarm probabiliy.

\begin{figure}[ht!]
        \includegraphics[width=90mm, height=60mm]{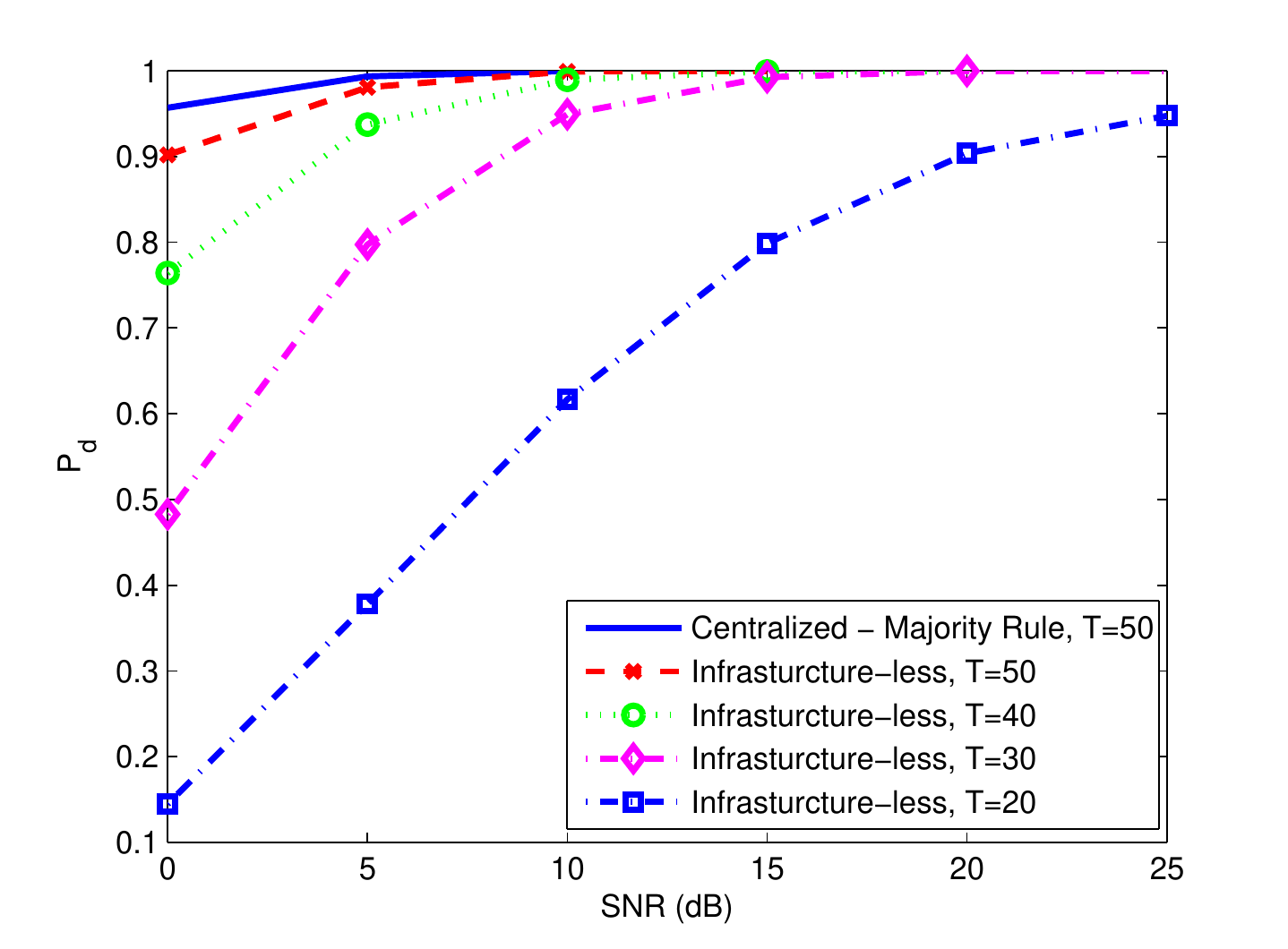}
       \caption{\small{Effect of number of measurements on probability of detection:  N = 200 channels, P = 4 users, M = 12 users, $\alpha$ = 2.}}
              \label{fig::Measurements}
    \end{figure}

Fig. \ref{fig::Measurements} shows the effect of the number of measurements on the detection performance. It is clearly seen that increasing the number of measurements will enhance both the detection probability and the false-alarm probabiliy.

\begin{figure}[ht!]
        \includegraphics[width=90mm, height=60mm]{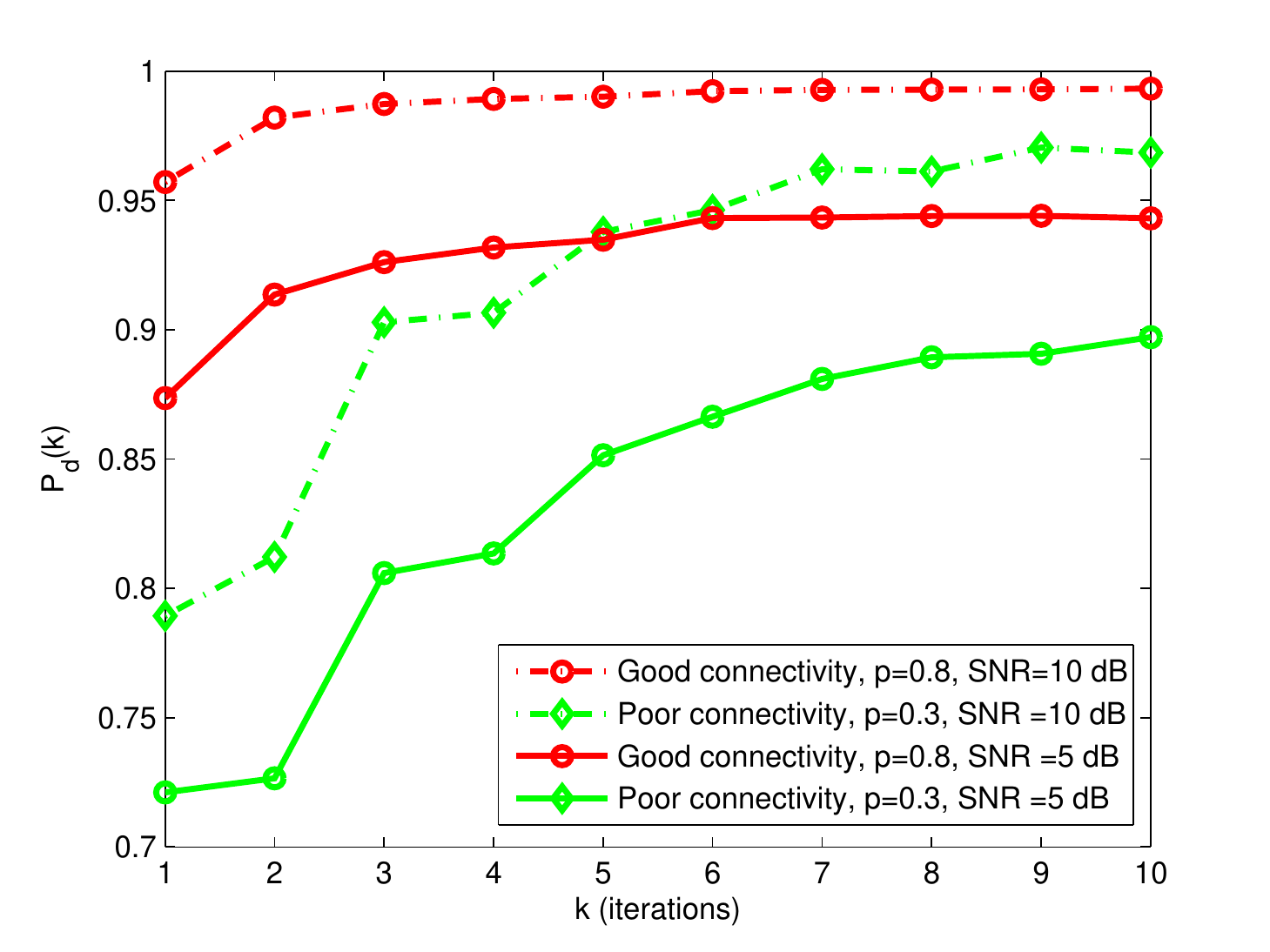}
        \caption{\small{The convergence of consensus algorithm in terms probability of detection: $N = 200$ channels, $T = 40$ measurements, $P = 4$ users, $M = 12$ users, $\alpha$ = 2.}}
                \label{fig::Convergence}
     \end{figure}

To show the effect of connectivity on the consensus algorithm, Fig. \ref{fig::Convergence} shows the probability of detection as a function of the number of algorithm iterations $K$ for $M = 12$ users. We consider two different scenarios of connectivity conditions between the SU nodes: 1) poor network connectivity ($p=0.3$), 2) good network connectivity ($p=0.8$) for two different values of SNR $5$ and $10$ dB. 

\begin{figure}[ht!]
        \includegraphics[scale=0.6]{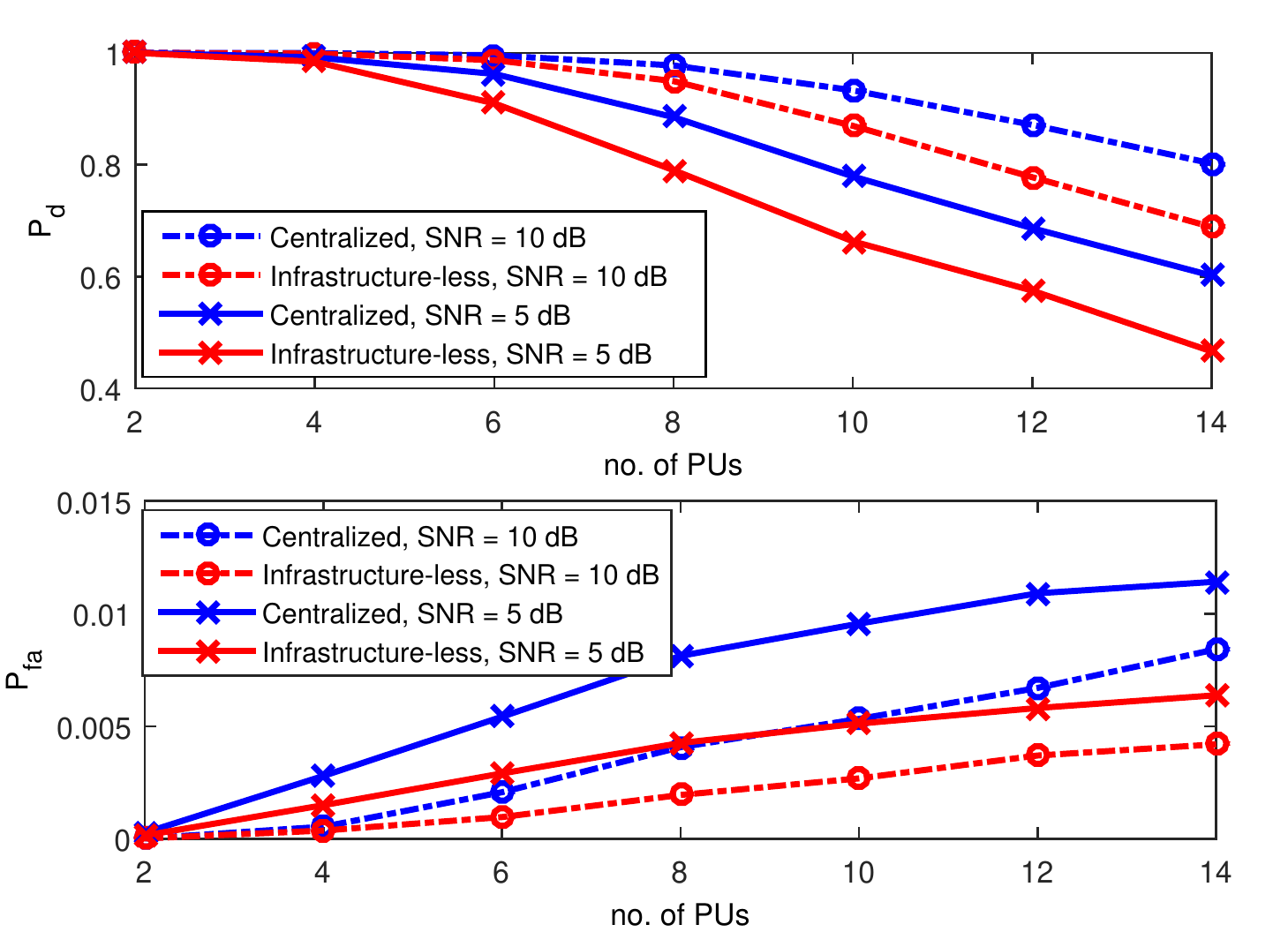}
        \caption{\small{Effect of PUs number : $N = 200$ channels, $T = 50$ measurements, $M = 12$ users, $\alpha = 2$, $p=0.8$.}}
        \label{fig::pu}
     \end{figure}

Fig. \ref{fig::pu} shows the performance degradation as the number of PUs increases. This indicates that, for a fixed number of measurements $T$, increasing the number of PUs will degrade the performance of detection. In other words, the sparsity of occupancy vector will reduce, and hence, the recovery for the occupancy vector will degrade (as the sparsity constraint becomes less valid as we increase the number of PUs).  

%
%

\section{Conclusion}

In this paper, we propose a distributed detection framework utilizing compressive sensing for infrastructure-less CRNs, which allows each SU to exchange its decisions with its neighbors at each iteration of a consensus algorithm. We have shown that an approximate consensus is reached after a sufficient number of iterations, depending on the link quality and the number of measurements at each node. Simulations results are used to study the different effects of the network parameters, like the link quality, the number of measurements, and the number of PUs, on the detection performance of the proposed algorithm.

\nocite{*}
\bibliographystyle{IEEEtran}
\bibliography{mybibfile}

\end{document}